\documentclass[useAMS,usenatbib]{mn2e}
 \usepackage{times,graphicx,amssymb,amsmath,natbib}

\newcommand{\be}{\begin{equation}}
\newcommand{\e}{\end{equation}}
\newcommand{\bear}{\begin{eqnarray}}
\newcommand{\ear}{\end{eqnarray}}

\newcommand{\f}{\frac}
\newcommand{\de}{{\rm d}}

\def\apjl{ApJL}
\def\apjs{ApJS}
\def\aap{A\&A}

\begin{document}

\title[Cosmic reionization after Planck]
{Cosmic reionization after Planck}
\author[Mitra, Choudhury \& Ferrara]
{Sourav Mitra$^1$\thanks{E-mail: hisourav@gmail.com},~
T. Roy Choudhury$^2$\thanks{E-mail: tirth@ncra.tifr.res.in}~
and
Andrea Ferrara$^3$\thanks{E-mail: andrea.ferrara@sns.it}\\
$^1$University of the Western Cape, Bellville, Cape Town 7535, South Africa\\
$^2$National Centre for Radio Astrophysics, TIFR, Post Bag 3, Ganeshkhind, Pune 411007, India\\
$^3$Scuola Normale Superiore, Piazza dei Cavalieri 7, 56126 Pisa, Italy
} 

\maketitle

\date{\today}

\begin{abstract}
Cosmic reionization holds the key to understand structure formation in the Universe,
and can inform us about the properties of the first sources, as their star formation
efficiency and escape fraction of ionizing photons. By combining the recent release
of Planck electron scattering optical depth data with observations of high-redshift
quasar absorption spectra, we obtain strong constraints on viable reionization
histories. We show that inclusion of Planck data favors a reionization scenario with
a single stellar population. The mean $x_{\rm HI}$ drops from $\sim0.8$ at $z=10.6$
to $\sim10^{-4}$ at $z=5.8$ and reionization is completed around $5.8\lesssim z\lesssim8.5$
(2-$\sigma$), thus indicating a significant reduction in contributions to reionization
from high redshift sources. We can put independent constraints on the escape fraction
$f_{\rm esc}$ of ionizing photons by incorporating the high-redshift galaxy luminosity
function data into our analysis. We find a non-evolving $f_{\rm esc}$ of
$\sim10\%$ in the redshift range $z=6-9$.
\end{abstract}

\begin{keywords}
dark ages, reionization, first stars -- intergalactic medium -- cosmology:
theory -- large-scale structure of Universe.
\end{keywords}

\section{Introduction}

Cosmic reionization is one of the key events in the history of Universe. Most of the
available constraints on the epoch of reionization (EoR) come from the observations
of the CMB by the Wilkinson Microwave Anisotropy Probe (WMAP) satellite
\citep{Komatsu:2010fb,2013ApJS..208...19H,2013ApJS..208...20B}, and from high redshift
QSOs \citep{Becker:2001ee,2003AJ....126....1W,2006AJ....132..117F}. The recent nine-year
WMAP observations provide the value of integrated Thomson scattering optical depth
$\tau_{\rm el}=0.089\pm0.014$ \citep{2013ApJS..208...19H}. This in turn corresponds to
an instantaneous reionization taking place at redshift $z_{\rm reion}=10.6\pm1.1$,
indicating a strong need for sources of reionization at $z\gtrsim10$. However, improved
measurements from three-year Planck mission suggest a lower value, $\tau_{\rm el}=0.066\pm0.012$,
\citep{2015arXiv150201589P} corresponding to $z_{\rm reion}=8.8^{+1.2}_{-1.1}$ and
therefore cuts down the demand for reionization sources beyond redshift $z=10$
\citep{2015ApJ...802L..19R,2015arXiv150308228B}. The resulting reionization histories
seem to explain the observations of Lyman-$\alpha$ emitters at $z\sim 7$
\citep{2015MNRAS.446..566M,2014arXiv1412.4790C} which were otherwise in tension with the
WMAP constraints. Although, these observations of cosmological data analysis are based
on the assumption that reionization is a sudden and instantaneous process, recent studies
\citep{BarkanaLoeb01,tirth06a,tirth06,tirth09,2010MNRAS.408...57P,mitra1,mitra2,2015arXiv150405601G}
clearly support a more extended process spanning the redshift range $6<z<15$. Several
semi-analytical models have been proposed with a combination of different observations
to put tighter limits on the reionization redshift and other quantities related to
reionization \citep{tirth05,2005ApJ...625....1W,2006MNRAS.370.1401G,2007MNRAS.379..253D,
2007MNRAS.377..285S,2008MNRAS.391...63I,mitra1,2011MNRAS.412.2781K,mitra2,2014ApJ...785...65C}.

Another observation set that could be used to check the consistency of such models is
Luminosity Function (LF) of high-$z$ ($6\lesssim z\lesssim10$) galaxies
\citep{2006NewAR..50..152B, 2012ApJ...745..110O,2012ApJ...760..108B,2014ApJ...786..108O,
2014arXiv1411.2976B,2014arXiv1412.1472M,2015ApJ...803...34B}. This procedure has to deal
with the yet poorly understood escape fraction of ionizing photons ($f_{\rm esc}$).
Despite of numerous impressive efforts in both observational and theoretical studies,
this quantity, as a function of galaxy mass and redshift, remains largely uncertain
\citep{2011ApJ...731...20F}. Available studies generally lead to a broad range and as well
as contradictory trends of $f_{\rm esc}$ on redshift. For example, \cite{2012ApJ...758...93F}
estimated average $f_{\rm esc}$ to be $\sim30\%$ in order to get a fully ionized IGM at $z=6$.
\cite{2012MNRAS.423..862K} found a strong redshift evolution of escape fraction increasing
from $\sim4\%$ at $z=4$ to unity at higher redshifts in order to simultaneously satisfy
reionization and lower redshift Lyman-$\alpha$ forest constraints. Based on a robust
statistical analysis on full CMB spectrum and quasar data and using the observations of
high-$z$ galaxy LFs, we \citep{mitra3} derived that mean value of $f_{\rm esc}$ is moderately
increasing from $7\%$ at $z=6$ to $18\%$ at $z=8$. This increasing behavior of $f_{\rm esc}$
on redshift is somewhat similar to that obtained or assumed in
\cite{2006MNRAS.371L...1I,2010ApJ...710.1239R,2011arXiv1103.5226H,
2012ApJ...746..125H,2013MNRAS.431.2826F,2015MNRAS.447.2526F}. More recently,
using a high-resolution cosmological zoom-in simulation of galaxy formation,
\cite{2015arXiv150307880M} found a considerably lower ($<5\%$; much less than required by
reionization models) and a non-evolving escape fraction. This unchanging trend of $f_{\rm esc}$
is as well consistent with many other earlier results
\citep{2008ApJ...673L...1G,2011MNRAS.412..411Y}. A decreasing tendency of escape
fraction with redshift has also been reported in the literature \citep{2000ApJ...545...86W,
2014ApJ...788..121K}.

For all these reasons, here we revise our reionization models \citep{mitra3} in the light of
recently available Planck data and improved measurements of high-$z$ LFs\footnote{Throughout
this Letter, we assume a flat Universe with Planck cosmological parameters: $\Omega_m = 0.3089$,
$\Omega_{\Lambda} = 1 - \Omega_m$, $\Omega_b h^2 = 0.02230$, $h=0.6774$, $\sigma_8=0.8159$,
$ n_s=0.9667$ and $Y_{\rm P}=0.2453$ \citep{2015arXiv150201589P}.}.

\section{Data-constrained reionization model}
\label{sec:cfmodel}

Let us first summarize the main features of the semi-analytical model for inhomogeneous
reionization used in this analysis, which is based on \cite{tirth05} and \cite{tirth06}. 
The model tracks the ionization and thermal evolution of all hydrogen and helium regions
separately and self-consistently by adopting a {\it lognormal} probability distribution\footnote{
Another commonly employed form of the probability density function (PDF) is an updated 
version of \cite{2000ApJ...530....1M} fit \citep{2009MNRAS.394.1812P}. However we checked that,
the differences in these two PDFs are much smaller than the errors in the data considered here
and thus the constraints are unlikely to be affected.}
at low densities, changing to a {\it power-law} at high densities
\citep{tirth05}. The model considers the inhomogeneities in the IGM according to the
description given by MHR \citep{2000ApJ...530....1M}, in which once all the low-density
regions are ionized, reionization is said to be complete (see \citealt{tirth09}). 

The sources of reionization are assumed to be stars (metal-free PopIII and normal PopII) and quasars.
The contribution of quasars are incorporated here based on their observed luminosity function
at $z<6$ \citep{2007ApJ...654..731H} and they have insignificant effects on IGM at higher redshifts
(but also see \citealt{2015A&A...578A..83G,2015arXiv150707678M}).
Furthermore, the model calculates the suppression of star formation in low-mass haloes ({\it radiative
feedback}) through a Jeans mass prescription, which is computed self-consistently from the evolution
of the thermal properties of IGM through the minimum circular velocity of haloes that are able to cool
(\citealt{tirth05}; also comparable to the simulations by \citealt{2008MNRAS.390..920O,2013MNRAS.432.3340S}).
Our model also accounts for the {\it chemical feedback}
(PopIII $\rightarrow$ PopII  transition) using merger-tree based genetic approach
\citep{2006MNRAS.369..825S}.The model computes the production rate of ionizing photons in the IGM as 
 \be
  \dot{n}_{\rm ph}(z) = n_b N_{\rm ion} \f{\de f_{\rm coll}}{\de t}
 \e
where, $f_{\rm coll}$ is the collapsed fraction of dark matter halo, $n_b$ is the total baryonic
number density in the IGM and $N_{\rm ion}$, possibly a function of halo mass and
redshift, is the number of photons entering the IGM per baryon in stars. However, throughout
this work we assume $N_{\rm ion}$ to be independent of halo mass.
This quantity can be written as $N_{\rm ion} = \epsilon_* f_{\rm esc} N_{\gamma}$, where
$\epsilon_*$ is the star formation efficiency, and $N_{\gamma}$ is the specific number of photons
emitted per baryon in stars (\citealt{mitra3}):

\subsection{MCMC-PCA constraints from Planck data}
\label{subsec: pca_Planck} 

\begin{figure*}
\begin{center}
  \includegraphics[height=0.798\textwidth, angle=270]{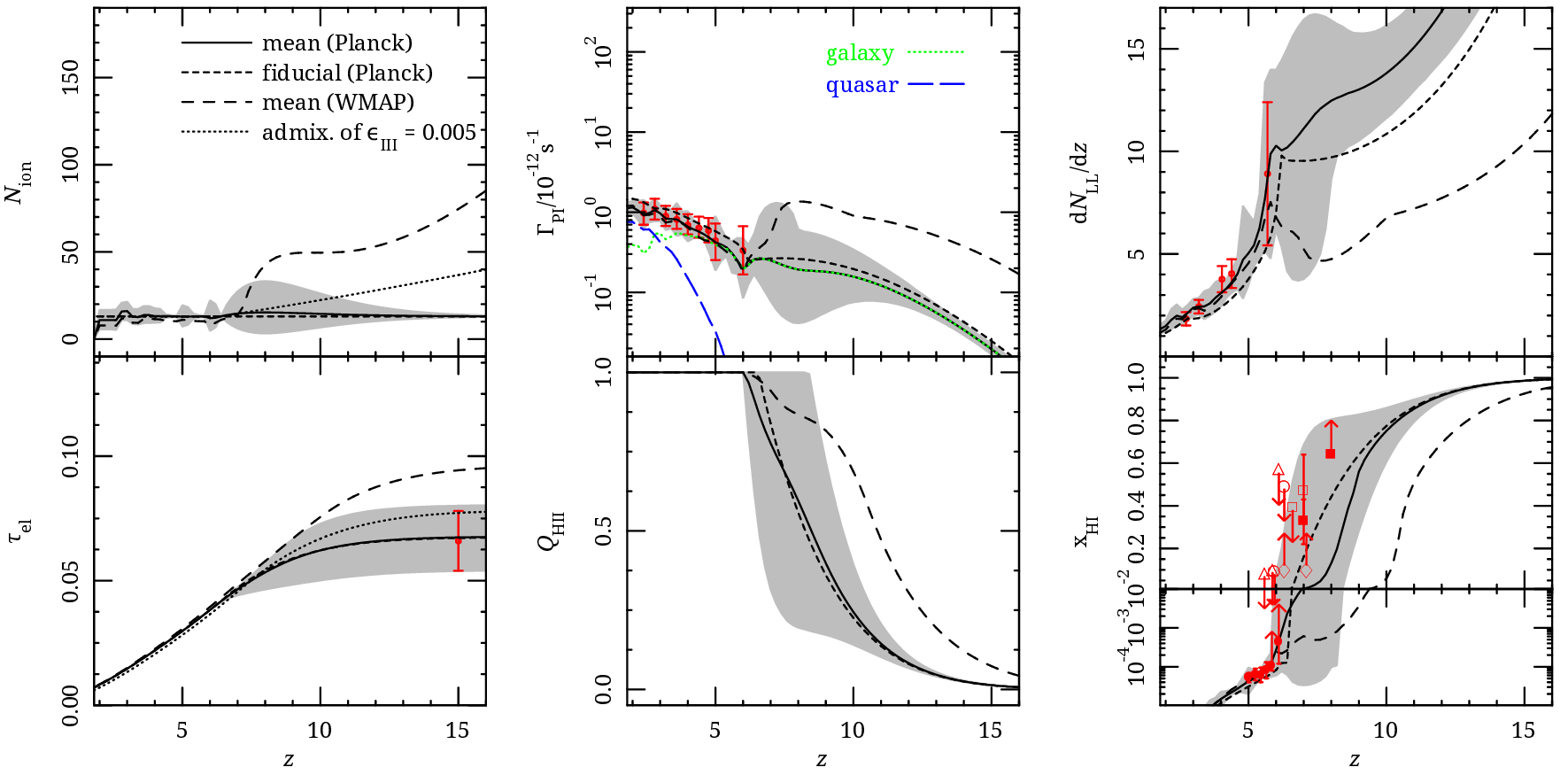}
  \caption{MCMC results: The mean value (solid lines) and its 2-$\sigma$ limits (shaded regions)
  for various quantities related to reionization obtained from our current analysis with Planck
  data. The fiducial model (short-dashed lines) and the model constrained using WMAP9
  $\tau_{\rm el}$ data (long-dashed lines) also shown for comparison. The red points with
  errorbars that we have used to constrain the model are taken from the observations of
  photoionization rates $\Gamma_{\rm PI}$ (\citealt{2007MNRAS.382..325B,2013MNRAS.436.1023B};
  {\it top-middle} panel), The redshift distribution of LLS $\de N_{\rm LL}/\de z$
  (\citealt{2010ApJ...721.1448S}; {\it top-right} panel) and the recent measurements of electron
  scattering optical depth $\tau_{\rm el}$ from Planck mission (\citealt{2015arXiv150201589P};
  {\it bottom-left} panel). We also show the observational limits on neutral hydrogen fraction
  $x_{\rm HI}(z)$ ({\it bottom-right} panel) from various measurements by
  \cite{2006AJ....132..117F} (filled circle), \cite{2015MNRAS.447..499M} (open triangle), 
  \cite{2006PASJ...58..485T,2013ApJ...774...26C} (open circle),
  \cite{2011MNRAS.416L..70B,2013MNRAS.428.3058S} (open diamond),
  \cite{2008ApJ...677...12O,2010ApJ...723..869O} (open square),
  \cite{2014ApJ...795...20S} (filled square).}
\label{fig:mcmc-PCA}
\end{center}
\end{figure*}

From the above model, we obtain the redshift evolution of $N_{\rm ion}(z)$ and other
quantities by performing a detailed likelihood estimation using the Principal Component
Analysis (PCA), following \cite{mitra1,mitra2}. We assume $N_{\rm ion}$ as an arbitrary
function of $z$ and decompose it into its principal components by constructing the Fisher
matrix from a fiducial model for $N_{\rm ion}$ using three different datasets: 
(i) measurements of photoionization rates $\Gamma_{\rm PI}$ in $2.4\leqslant z\leqslant6$
from \cite{2007MNRAS.382..325B} and \cite{2013MNRAS.436.1023B}
\footnote{The $\Gamma_{\rm PI}$ measurements of \cite{2007MNRAS.382..325B}
and \cite{2013MNRAS.436.1023B} somewhat depend on their choice of fiducial parameter set.
They have provided scaling relations to convert these measurements to any other choice of
parameters. While comparing with our models, we have applied appropriate scaling to those
for every parameter set considered. The $\Gamma_{\rm PI}$ data points with errorbars in
Fig. \ref{fig:mcmc-PCA} reflect the scaled measurements.};
(ii) redshift evolution of Lyman-limit systems (LLS), $\de N_{\rm LL}/\de z$ over a wide redshift
range ($0.36 < z < 6$) by \cite{2010ApJ...721.1448S}
\footnote{The reason for choosing \cite{2010ApJ...721.1448S} datasets over more recent measurements
on mean free path of ionizing photons by \cite{2014MNRAS.445.1745W} is that the former
data covers a wider redshift range, thus helping us to get tighter constraints
on reionization parameters.};
(iii) Thomson scattering optical depth
$\tau_{\rm el}$ using recent Planck data \citep{2015arXiv150201589P}.
We further impose fairly model-independent constraints on neutral hydrogen fraction
at $z\sim5-6$ from \cite{2015MNRAS.447..499M} as a prior to our model.
We choose the fiducial
model to be constant (no redshift dependence) as it matches the above-mentioned data points
quite accurately. When computing the ionizing radiation properties, we include only a single
stellar population (PopII) and neglect the contributions from PopIII sources (thus no
{\it chemical feedback} in the model) since such effects will be indirectly included in the
evolution of $N_{\rm ion}$ itself.

We then take the first $2-7$ PCA modes (eigen-modes of the Fisher matrix), having the largest
eigenvalues or smallest uncertainties, which satisfy a model-independent Akaike information
criteria (AIC; \citealt{2007MNRAS.377L..74L}) and finally perform
the Markov Chain Monte Carlo (MCMC) analysis\footnote{All cosmological parameters are fixed at
their best-fit Planck value.} using those modes; for details, see \cite{mitra1,mitra2}. 
As the Planck collaboration has not published low-$\ell$ polarization data, we include only
Planck optical depth data in the analysis. However, if one may wish to include the full
CMB spectra into the analysis, one should also consider variations in $\sigma_8$ and $n_s$
to avoid their possible degeneracies with the reionization model parameters
\citep{2011PhRvD..84l3522P,mitra2}.

The MCMC constraints on the model are shown in Fig. \ref{fig:mcmc-PCA}. The fiducial model
is well inside the shaded regions for all redshift range. One can see that, overall our model
predictions match the observed data points quite reasonably. We find that, all the quantities
are tightly constrained at $z\lesssim6$. This is expected as most of the observational
information considered in this work exists only at those redshifts. On the other hand,
a wide range of histories at $z> 6$ is still permitted by the data. The 2-$\sigma$ confidence
limits (C.L.) start to decrease at redshift $z\gtrsim13$ since
the components of the Fisher matrix are zero and there is no significant information
from the PCA modes beyond this point.

For comparison, we also show the model constrained by the WMAP9 $\tau_{\rm el}=0.089\pm0.014$
value and corresponding cosmological parameters \citep{2013ApJS..208...19H}. The mean evolution
of all the quantities for this model is almost identical to the Planck one at $z\lesssim6$;
at earlier epochs they start to differ, as expected from the different e.s. optical depth.
However, the mean model for WMAP9 lies within Planck's 2-$\sigma$ limits only up to $z\lesssim7$ or $8$.
We have further shown a typical admixture of PopIII contributions
with $\epsilon_{\rm PopIII}=0.005$ ($\epsilon\equiv\epsilon_* f_{\rm esc}$) to the $N_{\rm ion}$
(dotted black lines), which is the maximum limit that can be allowed by the Planck $\tau_{\rm el}$ data.
Whereas for the best-fit WMAP9 model, we get $\epsilon_{\rm PopIII}=0.014$.
The mean evolution of $N_{\rm ion}(z)$ for Planck closely follows the fiducial model, suggesting
that an non-evolving $N_{\rm ion}$ is well-permitted by the current data. This is one of the key
results from this work and deserves some more insight.

Unlike the WMAP9 model, the smaller value of $\tau_{\rm el}$  from Planck essentially releases the need
for high-redshift ionizing sources and allows the reionization to be completed from only
a single stellar population (PopII). The mean evolution of photoionization rates $\Gamma_{\rm PI}$
shows a mild increase at $z>6$, whereas
the WMAP model shows a relatively higher value of $\Gamma_{\rm PI}$ at early epochs as it
still allows the contributions coming from PopIII stars, which are able to produce large number
of ionizing photons. A similar trend is also found in the evolution of LLSs. Thus the future observations
on LLS around these epochs may able to further discriminate between these two models.
In the $\Gamma_{\rm PI}$ plot, we also show the relative contributions from quasars and galaxies
at different redshifts for the mean model by long-dashed (blue) and dotted (green) lines respectively.
From the evolution of volume filling factor $Q_{\rm HII}$ for ${\rm HII}$ regions, one can see that reionization is almost
completed ($Q_{\rm HII}\sim1$) around $5.8\lesssim z\lesssim8.5$ (2-$\sigma$ limits) for Planck model.
The mean ionized fraction evolves rapidly in $5.8\lesssim z\lesssim10.6$, in good agreement with other
recent works \citep{2015ApJ...799..177G,2015ApJ...802L..19R},
whereas the mean WMAP model favors a relatively gradual or extended reionization (spanning
$5.8\lesssim z\lesssim13$). This is reflected in the evolution of the neutral hydrogen
fraction $x_{\rm HI}$: the mean value for Planck model goes from
$x_{\rm HI}\sim0.8$ to $x_{\rm HI}\sim10^{-4}$ between $z=10.6$ ($z=13$ for WMAP model) and $z=5.8$.
Overall our model prediction for $x_{\rm HI}(z)$ is consistent with various observational limits
(see the caption in the Fig. \ref{fig:mcmc-PCA} for references).

\section{UV Luminosity Function}
\label{sec:uvluminosity}

\begin{figure*}
\includegraphics[height=0.76\textwidth, width=0.172\textheight, angle=270]{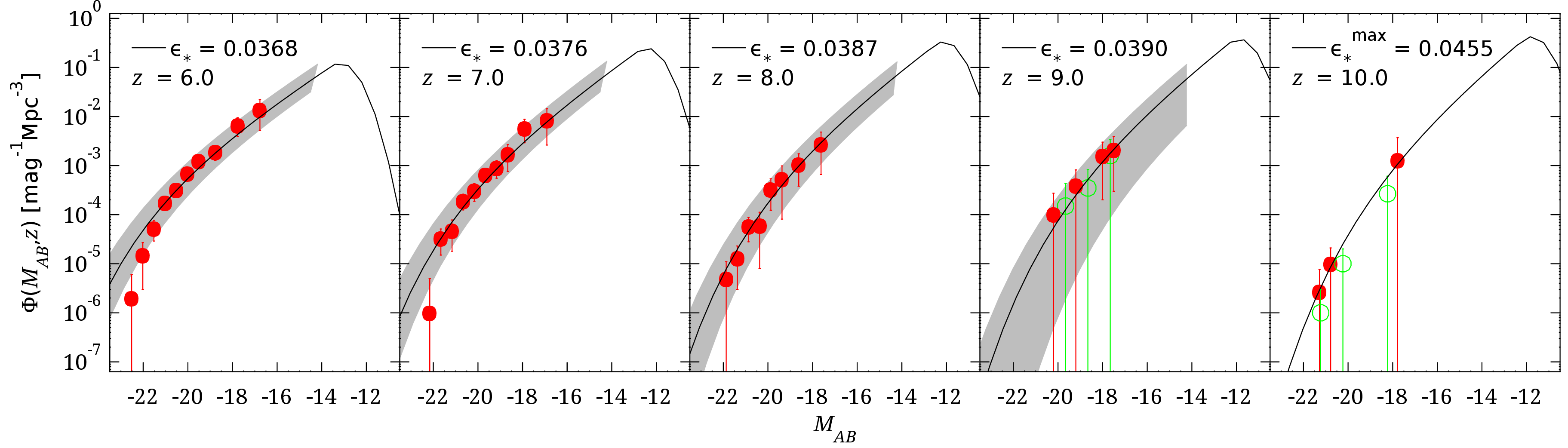}
\includegraphics[height=0.23\textwidth, width=0.165\textheight, angle=270]{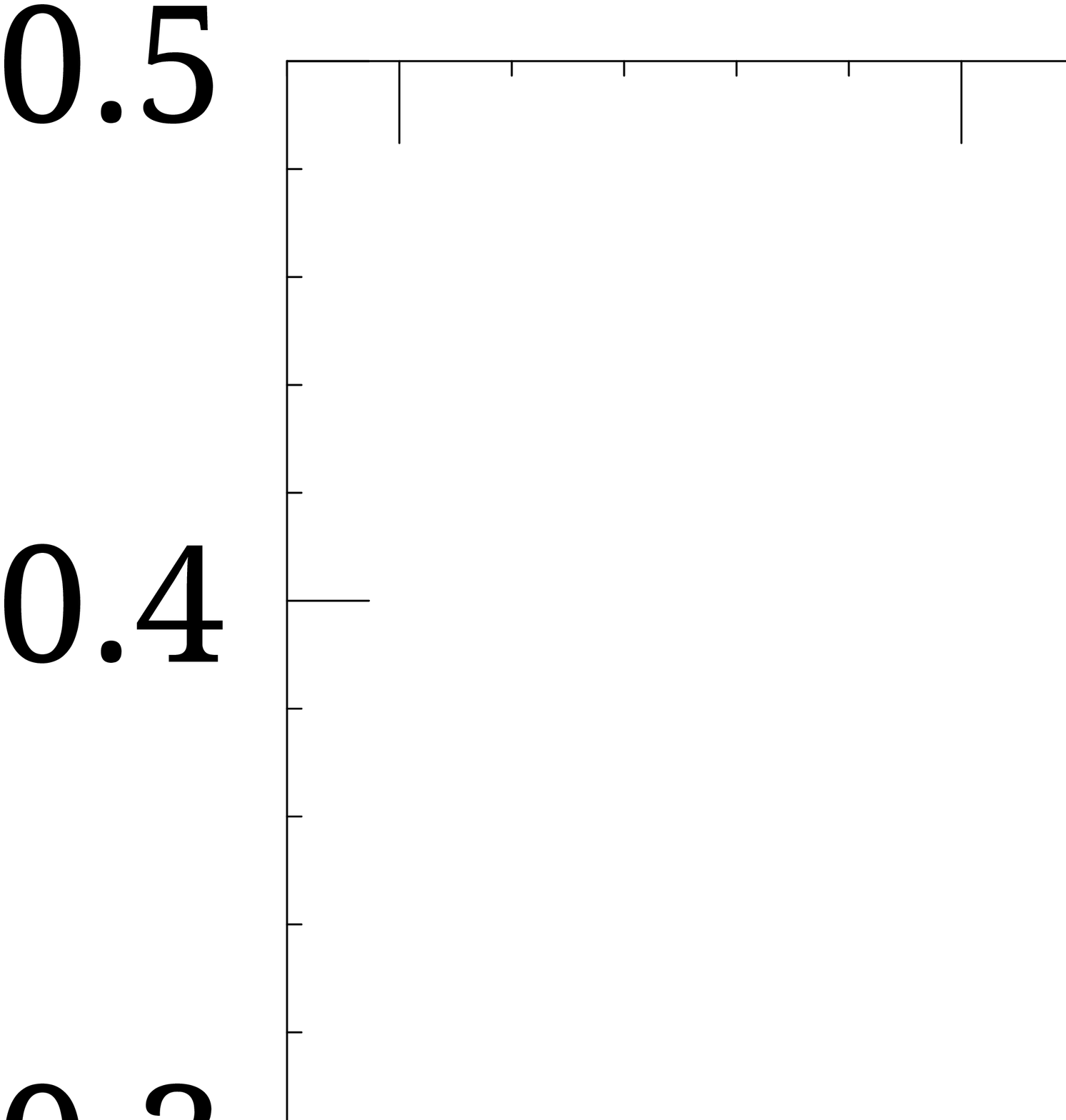}
\caption{{\it Left} panels: Evolution of luminosity function from our model for best-fit $\epsilon_*$ (black curve) 
and 2-$\sigma$ limits (shaded region) at $z=6-10$. Data (red) points with 2-$\sigma$ errors
are: \cite{2015ApJ...803...34B} (for $z=6,7,8$), combined datasets from \cite{2013MNRAS.432.2696M}
and \cite{2014arXiv1412.1472M} for $z=9$ and improved data from
\cite{2014ApJ...786..108O} for $z=10$. For completeness, we also show data from
\cite{2013ApJ...773...75O} for $z=9$ and \cite{2015ApJ...803...34B} for $z=10$
(not included in our analysis, open green points).
{\it Right} panel: Redshift evolution of the escape fraction $f_{\rm esc}$ with 2-$\sigma$ errors.
The $z=10$ point shows the lower limit on $f_{\rm esc}$ at that redshift.}
\label{fig:lumfunc}
\end{figure*}

From the above data-constrained reionization models, we now derive the high-$z$ LFs, following \citealt{mitra3}.
The luminosity at $1500$ $\mathring{\rm A}$ of a galaxy with mass $M$ and age $\Delta t$
($=t_{z}-t_{z'}$; time elapsed between the redshift of formation $z'$ and redshift of observation
$z$) can be written as \citep{2007MNRAS.377..285S,2011MNRAS.412.2781K}:
\be
 L_{1500} (M, \Delta t) = \epsilon_* \left(\frac{\Omega_b}{\Omega_m}\right)M l_{1500}(\Delta t)
 \label{eq:epsilon}
\e
where, $\epsilon_*$ is the star-forming efficiency of PopII stars only as we will restrict ourselves
to a single stellar population throughout this analysis. The template specific luminosity $l_{1500}$
is computed from stellar population synthesis model of \cite{2003MNRAS.344.1000B} for PopII stars
having different metallicities in the range $Z=0.0001-0.05$. Here we have incorporated the
mass-metallicity relation given by \cite{dayal1} and \cite{dayal2} and take the appropriate model
with that metallicity. The luminosity is then converted to standard absolute AB magnitude system
\citep{1983ApJ...266..713O,2011MNRAS.412.2781K}, and finally we obtain the LF, $\Phi (M_{AB},z)$, from
\be
 \Phi (M_{AB},z) = \frac{\de n}{\de M_{AB}} = \frac{\de n}{\de L_{1500}}\frac{\de L_{1500}}{\de M_{AB}}
 \label{eq:LF}
\e
where $\de n/\de L_{1500}$ is the comoving number of objects having luminosity within
$[L_{1500}, L_{1500} + \de L_{1500}]$ at a redshift $z$. This quantity can be calculated from the
formation rate of haloes using our reionization model \citep{tirth07}.

Now, we vary $\epsilon_*$ in eq. \ref{eq:epsilon} as a free parameter and match the observed LFs
with our model predictions computed using eq. \ref{eq:LF}. In Fig. \ref{fig:lumfunc} ({\it left} panels),
we show our results for the best-fit $\epsilon_*$ with $95\%$ C.L. for redshifts
$z=6-10$. As the observations at $z=10$ are still scant, we are able to determine only an upper limit
of $\epsilon_*$. The best-fit $\epsilon_*$ remains almost constant ($\sim4\%$) for all redshift range.
The number of galaxies start to decrease at fainter magnitudes producing a ``knee''-like shape
in the LFs which shifts towards the brighter ends at lower redshifts ($M_{AB}\sim-12$ at $z=10$ to
$M_{AB}\sim-14$ at $z=6$). This is due to the radiative feedback implemented in our model where
the star formation becomes suppressed.

Overall, the match between data and model predictions is quite impressive for all redshifts considered
here. The uncertainties (shaded regions) are larger at high-$z$ and at the bright end of LFs. For lower
redshifts ($z\leqslant7$), although our model can match the fainter end of the LF accurately, it slightly
over-predicts the bright end. This general tendency has already been addressed in several
recent works (\citealt{2014ApJ...785...65C,2014MNRAS.445.2545D,2014arXiv1411.2976B} and the references
therein). In particular, \cite{2014ApJ...785...65C} argued that it can be resolved by taking the
dust obscuration into account, which we are neglecting here. As the dust extinction was insignificant
at earlier times, we are getting a good match for the brighter end at $z>7$. Alternatively, the surveyed
volumes might be too small to catch the brightest, rare objects (for a discussion, see
\citealt{2014arXiv1411.2976B,2014MNRAS.445.2545D}). However, it is still unclear whether this discrepancy
arises from neglecting dust or halo mass quenching \citep{2010ApJ...721..193P}, or it is due to a
mass-dependent $\epsilon_*$. Thus it would be interesting to take those effects in our model which
we prefer to leave for future work.

\begin{table}
\centering
\begin{tabular}{c|c|c}
Redshift & best-fit $\epsilon_*$ [2-$\sigma$ limits] & best-fit $f_{\rm esc}$ [2-$\sigma$limits]\\
\hline
$z=6$ & $0.0368$ $[0.0189,0.0743]$ & $0.1018$ $[0.0186,0.2018]$\\
$z=7$ & $0.0376$ $[0.0194,0.0773]$ & $0.1215$ $[0.0247,0.2424]$\\
$z=8$ & $0.0387$ $[0.0170,0.0764]$ & $0.1234$ $[0.0309,0.2839]$\\
$z=9$ & $0.0390$ $[0.0050,0.0785]$ & $0.1202$ $[0.0276,0.2838]$\\
$z=10$& $< 0.0455$ & $> 0.0996$\\
\hline
\end{tabular}
\caption{Best-fit values and 2-$\sigma$ limits of star-forming efficiency $\epsilon_*$ and the
escape fraction $f_{\rm esc}$ obtained from the LF matching at different redshifts $z=6-10$.
At $z=10$, we only get an upper limit of $\epsilon_*$ and a corresponding lower limit of $f_{\rm esc}$.} 
\label{tab:escfrac}
\end{table}

\subsection{Escape fraction evolution}
\label{subsec:escfrac}
As a final step, having fixed $\epsilon_*$ for different redshifts, we can derive
limits for $f_{\rm esc}$ using the reionization constraints on the evolution of
$N_{\rm ion}$ (Sec. \ref{subsec: pca_Planck}) with $N_{\gamma} = 3200$ as appropriate
for the PopII Salpeter IMF assumed here. The uncertainties in $f_{\rm esc}$ can also
be calculated using the quadrature method \citep{mitra3}. We show our resulting
$f_{\rm esc}$ in Table \ref{tab:escfrac} and the {\it right} panel of Fig. \ref{fig:lumfunc}. We find
almost non-evolving (constant) $f_{\rm esc}$ of $\sim10\%$ (best-fit) in the redshift range $z=6-9$.
For $z=10$, we get a lower limit of $10\%$ which is a very strong
constraint given the uncertainties present in high-$z$ observations.


\section{Conclusions}
\label{sec:conclusions}
Over the past few years, several numerical and analytical approaches have tried to constrain 
reionization scenarios by using CMB WMAP observations and QSOs. In particular, in our earlier
works \citep{mitra1,mitra2}, we proposed a detailed semi-analytical modeling of hydrogen reionization using 
the observations for photoionization rates, redshift evolution of LLS and CMB and succeeded to produce a good
match with a variety of other relevant datasets. We further tested the model against the observations of luminosity
functions from high-redshift galaxies \citep{mitra3}. In this Letter, we extend those works by taking the 
recently published $\tau_{\rm el}$ data from \cite{2015arXiv150201589P}
and various new results from the observations of galaxy LFs at $6\leqslant z\leqslant10$.

\begin{itemize}
 \item We find that, contrary to WMAP data, a constant/non-evolving $N_{\rm ion}$ is now allowed by the
 Planck data. This immediately tells us that, reionization with a single stellar population (PopII) or
 non-evolving IMF is possible, i.e. the impact of PopIII stars on reionization \citep{2013MNRAS.429L..94P}
 is likely to be negligible.  \item According to Planck data, reionization proceeds quickly from $z\approx10.6$ to
 $z\approx5.8$ as the mean $x_{\rm HI}$ drops from $0.8$ to $10^{-4}$ within these
 epochs. We find that reionization is almost completed around $5.8\lesssim z\lesssim8.5$ (2-$\sigma$ C.L.).
 However, the model with WMAP data seems to favor an extended reionization starting as early
 as $z\approx13$.
 Thus, the inclusion of Planck data in turn indicates that most of the reionization activity occurs at
 $z\lesssim10$ \citep{2015ApJ...802L..19R,2015arXiv150308228B}.
 \item From the match between the observed high-redshift LFs and our model predictions, we find that the
 best-fit values for both $\epsilon_*$ and $f_{\rm esc}$ remains somewhat constant with redshifts:
 $\epsilon_*$ at $\sim4\%$ and $f_{\rm esc}\sim10\%$ for $z=6-9$
 \citep{2008ApJ...673L...1G,2011MNRAS.412..411Y,2015arXiv150307880M}. We have also obtained the tightest
 constraint available to our knowledge on on $f_{\rm esc}$ ($>10\%$) at $z=10$.
\end{itemize}

Although we have focused on high-redshift LFs, one can apply the same method for the evolution in
lower redshift range $3\leqslant z\leqslant5$. As the dust extinction becomes significant at those
redshifts, one has to take that into account. Moreover, the addition of dust and/or a mass-dependent
$\epsilon_*$ may resolve the problem of overproducing the brighter end of LFs, as stated
earlier. We defer these issues to future work. Also, the inclusion of {\it full} CMB datasets from
Planck into our analysis can be helpful for ruling out some of the current reionization scenarios.
Unfortunately, the recent Planck data release does not include the polarization data in their likelihood; instead,
they rely on the WMAP polarization likelihood at low multipoles \citep{2015arXiv150201589P}. As most of
the constraints at $z>6$ related to reionization models come from polarization data \citep{mitra2},
we postpone such analysis to the next Planck data release.
\bibliography{reionization-smitra}

\begin{thebibliography}{}

\bibitem[\protect\citeauthoryear{{Barkana} \& {Loeb}}{{Barkana} \&
  {Loeb}}{2001}]{BarkanaLoeb01}
{Barkana} R.,  {Loeb} A.,  2001, \physrep, 349, 125

\bibitem[\protect\citeauthoryear{{Becker} \& {Bolton}}{{Becker} \&
  {Bolton}}{2013}]{2013MNRAS.436.1023B}
{Becker} G.~D.,  {Bolton} J.~S.,  2013, \mnras, 436, 1023

\bibitem[\protect\citeauthoryear{Becker et~al.,}{Becker
  et~al.}{2001}]{Becker:2001ee}
Becker R.~H.,  et~al., 2001, Astron.J., 122, 2850

\bibitem[\protect\citeauthoryear{{Bennett} et~al.,}{{Bennett}
  et~al.}{2013}]{2013ApJS..208...20B}
{Bennett} C.~L.,  et~al., 2013, \apjs, 208, 20

\bibitem[\protect\citeauthoryear{{Bolton} et~al.,}{{Bolton}
  et~al.}{2011}]{2011MNRAS.416L..70B}
{Bolton} J.~S.,  et~al., 2011, \mnras, 416, L70

\bibitem[\protect\citeauthoryear{{Bolton} \& {Haehnelt}}{{Bolton} \&
  {Haehnelt}}{2007}]{2007MNRAS.382..325B}
{Bolton} J.~S.,  {Haehnelt} M.~G.,  2007, \mnras, 382, 325

\bibitem[\protect\citeauthoryear{{Bouwens} \& {Illingworth}}{{Bouwens} \&
  {Illingworth}}{2006}]{2006NewAR..50..152B}
{Bouwens} R.,  {Illingworth} G.,  2006, \nar, 50, 152

\bibitem[\protect\citeauthoryear{{Bouwens} et~al.,}{{Bouwens}
  et~al.}{2015a}]{2015arXiv150308228B}
{Bouwens} R.~J.,  et~al., 2015a, arXiv:1503.08228

\bibitem[\protect\citeauthoryear{{Bouwens} et~al.,}{{Bouwens}
  et~al.}{2015b}]{2015ApJ...803...34B}
{Bouwens} R.~J.,  et~al., 2015b, \apj, 803, 34

\bibitem[\protect\citeauthoryear{{Bowler} et~al.,}{{Bowler}
  et~al.}{2014}]{2014arXiv1411.2976B}
{Bowler} R.~A.~A.,  et~al., 2014, arXiv:1411.2976

\bibitem[\protect\citeauthoryear{{Bradley} et~al.,}{{Bradley}
  et~al.}{2012}]{2012ApJ...760..108B}
{Bradley} L.~D.,  et~al., 2012, \apj, 760, 108

\bibitem[\protect\citeauthoryear{{Bruzual} \& {Charlot}}{{Bruzual} \&
  {Charlot}}{2003}]{2003MNRAS.344.1000B}
{Bruzual} G.,  {Charlot} S.,  2003, \mnras, 344, 1000

\bibitem[\protect\citeauthoryear{{Cai}, {Lapi}, {Bressan}, {De Zotti},
  {Negrello} \& {Danese}}{{Cai} et~al.}{2014}]{2014ApJ...785...65C}
{Cai} Z.-Y.,  {Lapi} A.,  {Bressan} A.,  {De Zotti} G.,  {Negrello} M.,
  {Danese} L.,  2014, \apj, 785, 65

\bibitem[\protect\citeauthoryear{{Chornock} et~al.,}{{Chornock}
  et~al.}{2013}]{2013ApJ...774...26C}
{Chornock} R.,  et~al., 2013, \apj, 774, 26

\bibitem[\protect\citeauthoryear{{Choudhury}}{{Choudhury}}{2009}]{tirth09}
{Choudhury} T.~R.,  2009, Current Science, 97, 841

\bibitem[\protect\citeauthoryear{{Choudhury} \& {Ferrara}}{{Choudhury} \&
  {Ferrara}}{2005}]{tirth05}
{Choudhury} T.~R.,  {Ferrara} A.,  2005, \mnras, 361, 577

\bibitem[\protect\citeauthoryear{{Choudhury} \& {Ferrara}}{{Choudhury} \&
  {Ferrara}}{2006a}]{tirth06a}
{Choudhury} T.~R.,  {Ferrara} A.,  2006a, arXiv:astro-ph/0603149

\bibitem[\protect\citeauthoryear{{Choudhury} \& {Ferrara}}{{Choudhury} \&
  {Ferrara}}{2006b}]{tirth06}
{Choudhury} T.~R.,  {Ferrara} A.,  2006b, \mnras, 371, L55

\bibitem[\protect\citeauthoryear{{Choudhury} \& {Ferrara}}{{Choudhury} \&
  {Ferrara}}{2007}]{tirth07}
{Choudhury} T.~R.,  {Ferrara} A.,  2007, \mnras, 380, L6

\bibitem[\protect\citeauthoryear{{Choudhury}, {Puchwein}, {Haehnelt} \&
  {Bolton}}{{Choudhury} et~al.}{2014}]{2014arXiv1412.4790C}
{Choudhury} T.~R.,  {Puchwein} E.,  {Haehnelt} M.~G.,    {Bolton} J.~S.,  2014,
  arXiv:1412.4790

\bibitem[\protect\citeauthoryear{{Dayal}, {Ferrara}, {Dunlop} \&
  {Pacucci}}{{Dayal} et~al.}{2014}]{2014MNRAS.445.2545D}
{Dayal} P.,  {Ferrara} A.,  {Dunlop} J.~S.,    {Pacucci} F.,  2014, \mnras,
  445, 2545

\bibitem[\protect\citeauthoryear{{Dayal}, {Ferrara} \& {Saro}}{{Dayal}
  et~al.}{2010}]{dayal2}
{Dayal} P.,  {Ferrara} A.,    {Saro} A.,  2010, \mnras, 402, 1449

\bibitem[\protect\citeauthoryear{{Dayal}, {Ferrara}, {Saro}, {Salvaterra},
  {Borgani} \& {Tornatore}}{{Dayal} et~al.}{2009}]{dayal1}
{Dayal} P.,  {Ferrara} A.,  {Saro} A.,  {Salvaterra} R.,  {Borgani} S.,
  {Tornatore} L.,  2009, \mnras, 400, 2000

\bibitem[\protect\citeauthoryear{{Dijkstra}, {Wyithe} \& {Haiman}}{{Dijkstra}
  et~al.}{2007}]{2007MNRAS.379..253D}
{Dijkstra} M.,  {Wyithe} J.~S.~B.,    {Haiman} Z.,  2007, \mnras, 379, 253

\bibitem[\protect\citeauthoryear{{Fan} et~al.,}{{Fan}
  et~al.}{2006}]{2006AJ....132..117F}
{Fan} X.,  et~al., 2006, \aj, 132, 117

\bibitem[\protect\citeauthoryear{{Fernandez} \& {Shull}}{{Fernandez} \&
  {Shull}}{2011}]{2011ApJ...731...20F}
{Fernandez} E.~R.,  {Shull} J.~M.,  2011, \apj, 731, 20

\bibitem[\protect\citeauthoryear{{Ferrara} \& {Loeb}}{{Ferrara} \&
  {Loeb}}{2013}]{2013MNRAS.431.2826F}
{Ferrara} A.,  {Loeb} A.,  2013, \mnras, 431, 2826

\bibitem[\protect\citeauthoryear{{Finkelstein} et~al.,}{{Finkelstein}
  et~al.}{2012}]{2012ApJ...758...93F}
{Finkelstein} S.~L.,  et~al., 2012, \apj, 758, 93

\bibitem[\protect\citeauthoryear{{Finlator}, {Thompson}, {Huang}, {Dav{\'e}},
  {Zackrisson} \& {Oppenheimer}}{{Finlator} et~al.}{2015}]{2015MNRAS.447.2526F}
{Finlator} K.,  {Thompson} R.,  {Huang} S.,  {Dav{\'e}} R.,  {Zackrisson} E.,
   {Oppenheimer} B.~D.,  2015, \mnras, 447, 2526

\bibitem[\protect\citeauthoryear{{Gallerani}, {Choudhury} \&
  {Ferrara}}{{Gallerani} et~al.}{2006}]{2006MNRAS.370.1401G}
{Gallerani} S.,  {Choudhury} T.~R.,    {Ferrara} A.,  2006, \mnras, 370, 1401

\bibitem[\protect\citeauthoryear{{George} et~al.,}{{George}
  et~al.}{2015}]{2015ApJ...799..177G}
{George} E.~M.,  et~al., 2015, \apj, 799, 177

\bibitem[\protect\citeauthoryear{{Ghara}, {Datta} \& {Choudhury}}{{Ghara}
  et~al.}{2015}]{2015arXiv150405601G}
{Ghara} R.,  {Datta} K.~K.,    {Choudhury} T.~R.,  2015, arXiv:1504.05601

\bibitem[\protect\citeauthoryear{{Giallongo} et~al.,}{{Giallongo}
  et~al.}{2015}]{2015A&A...578A..83G}
{Giallongo} E.,  et~al., 2015, \aap, 578, A83

\bibitem[\protect\citeauthoryear{{Gnedin}}{{Gnedin}}{2008}]{2008ApJ...673L...1%
G}
{Gnedin} N.~Y.,  2008, \apjl, 673, L1

\bibitem[\protect\citeauthoryear{{Haardt} \& {Madau}}{{Haardt} \&
  {Madau}}{2011}]{2011arXiv1103.5226H}
{Haardt} F.,  {Madau} P.,  2011, arXiv:1103.5226

\bibitem[\protect\citeauthoryear{{Haardt} \& {Madau}}{{Haardt} \&
  {Madau}}{2012}]{2012ApJ...746..125H}
{Haardt} F.,  {Madau} P.,  2012, \apj, 746, 125

\bibitem[\protect\citeauthoryear{{Hinshaw} et~al.,}{{Hinshaw}
  et~al.}{2013}]{2013ApJS..208...19H}
{Hinshaw} G.,  et~al., 2013, \apjs, 208, 19

\bibitem[\protect\citeauthoryear{{Hopkins}, {Richards} \&
  {Hernquist}}{{Hopkins} et~al.}{2007}]{2007ApJ...654..731H}
{Hopkins} P.~F.,  {Richards} G.~T.,    {Hernquist} L.,  2007, \apj, 654, 731

\bibitem[\protect\citeauthoryear{{Iliev}, {Shapiro}, {McDonald}, {Mellema} \&
  {Pen}}{{Iliev} et~al.}{2008}]{2008MNRAS.391...63I}
{Iliev} I.~T.,  {Shapiro} P.~R.,  {McDonald} P.,  {Mellema} G.,    {Pen} U.-L.,
   2008, \mnras, 391, 63

\bibitem[\protect\citeauthoryear{{Inoue}, {Iwata} \& {Deharveng}}{{Inoue}
  et~al.}{2006}]{2006MNRAS.371L...1I}
{Inoue} A.~K.,  {Iwata} I.,    {Deharveng} J.-M.,  2006, \mnras, 371, L1

\bibitem[\protect\citeauthoryear{{Kimm} \& {Cen}}{{Kimm} \&
  {Cen}}{2014}]{2014ApJ...788..121K}
{Kimm} T.,  {Cen} R.,  2014, \apj, 788, 121

\bibitem[\protect\citeauthoryear{Komatsu et~al.,}{Komatsu
  et~al.}{2011}]{Komatsu:2010fb}
Komatsu E.,  et~al., 2011, Astrophys.J.Suppl., 192, 18

\bibitem[\protect\citeauthoryear{{Kuhlen} \& {Faucher-Gigu{\`e}re}}{{Kuhlen} \&
  {Faucher-Gigu{\`e}re}}{2012}]{2012MNRAS.423..862K}
{Kuhlen} M.,  {Faucher-Gigu{\`e}re} C.-A.,  2012, \mnras, 423, 862

\bibitem[\protect\citeauthoryear{{Kulkarni} \& {Choudhury}}{{Kulkarni} \&
  {Choudhury}}{2011}]{2011MNRAS.412.2781K}
{Kulkarni} G.,  {Choudhury} T.~R.,  2011, \mnras, 412, 2781

\bibitem[\protect\citeauthoryear{{Liddle}}{{Liddle}}{2007}]{2007MNRAS.377L..74%
L}
{Liddle} A.~R.,  2007, \mnras, 377, L74

\bibitem[\protect\citeauthoryear{{Ma}, {Kasen}, {Hopkins}, {Faucher-Giguere},
  {Quataert}, {Keres} \& {Murray}}{{Ma} et~al.}{2015}]{2015arXiv150307880M}
{Ma} X.,  {Kasen} D.,  {Hopkins} P.~F.,  {Faucher-Giguere} C.-A.,  {Quataert}
  E.,  {Keres} D.,    {Murray} N.,  2015, arXiv:1503.07880

\bibitem[\protect\citeauthoryear{{Madau} \& {Haardt}}{{Madau} \&
  {Haardt}}{2015}]{2015arXiv150707678M}
{Madau} P.,  {Haardt} F.,  2015, arXiv:1507.07678

\bibitem[\protect\citeauthoryear{{McGreer}, {Mesinger} \&
  {D'Odorico}}{{McGreer} et~al.}{2015}]{2015MNRAS.447..499M}
{McGreer} I.~D.,  {Mesinger} A.,    {D'Odorico} V.,  2015, \mnras, 447, 499

\bibitem[\protect\citeauthoryear{{McLeod}, {McLure}, {Dunlop}, {Robertson},
  {Ellis} \& {Targett}}{{McLeod} et~al.}{2014}]{2014arXiv1412.1472M}
{McLeod} D.~J.,  {McLure} R.~J.,  {Dunlop} J.~S.,  {Robertson} B.~E.,  {Ellis}
  R.~S.,    {Targett} T.~T.,  2014, arXiv:1412.1472

\bibitem[\protect\citeauthoryear{{McLure} et~al.,}{{McLure}
  et~al.}{2013}]{2013MNRAS.432.2696M}
{McLure} R.~J.,  et~al., 2013, \mnras, 432, 2696

\bibitem[\protect\citeauthoryear{{Mesinger}, {Aykutalp}, {Vanzella},
  {Pentericci}, {Ferrara} \& {Dijkstra}}{{Mesinger}
  et~al.}{2015}]{2015MNRAS.446..566M}
{Mesinger} A.,  {Aykutalp} A.,  {Vanzella} E.,  {Pentericci} L.,  {Ferrara} A.,
     {Dijkstra} M.,  2015, \mnras, 446, 566

\bibitem[\protect\citeauthoryear{{Miralda-Escud{\'e}}, {Haehnelt} \&
  {Rees}}{{Miralda-Escud{\'e}} et~al.}{2000}]{2000ApJ...530....1M}
{Miralda-Escud{\'e}} J.,  {Haehnelt} M.,    {Rees} M.~J.,  2000, \apj, 530, 1

\bibitem[\protect\citeauthoryear{{Mitra}, {Choudhury} \& {Ferrara}}{{Mitra}
  et~al.}{2011}]{mitra1}
{Mitra} S.,  {Choudhury} T.~R.,    {Ferrara} A.,  2011, \mnras, 413, 1569

\bibitem[\protect\citeauthoryear{{Mitra}, {Choudhury} \& {Ferrara}}{{Mitra}
  et~al.}{2012}]{mitra2}
{Mitra} S.,  {Choudhury} T.~R.,    {Ferrara} A.,  2012, \mnras, 419, 1480

\bibitem[\protect\citeauthoryear{{Mitra}, {Ferrara} \& {Choudhury}}{{Mitra}
  et~al.}{2013}]{mitra3}
{Mitra} S.,  {Ferrara} A.,    {Choudhury} T.~R.,  2013, \mnras, 428, L1

\bibitem[\protect\citeauthoryear{{Oesch} et~al.,}{{Oesch}
  et~al.}{2012}]{2012ApJ...745..110O}
{Oesch} P.~A.,  et~al., 2012, \apj, 745, 110

\bibitem[\protect\citeauthoryear{{Oesch} et~al.,}{{Oesch}
  et~al.}{2013}]{2013ApJ...773...75O}
{Oesch} P.~A.,  et~al., 2013, \apj, 773, 75

\bibitem[\protect\citeauthoryear{{Oesch} et~al.,}{{Oesch}
  et~al.}{2014}]{2014ApJ...786..108O}
{Oesch} P.~A.,  et~al., 2014, \apj, 786, 108

\bibitem[\protect\citeauthoryear{{Okamoto}, {Gao} \& {Theuns}}{{Okamoto}
  et~al.}{2008}]{2008MNRAS.390..920O}
{Okamoto} T.,  {Gao} L.,    {Theuns} T.,  2008, \mnras, 390, 920

\bibitem[\protect\citeauthoryear{{Oke} \& {Gunn}}{{Oke} \&
  {Gunn}}{1983}]{1983ApJ...266..713O}
{Oke} J.~B.,  {Gunn} J.~E.,  1983, \apj, 266, 713

\bibitem[\protect\citeauthoryear{{Ota} et~al.,}{{Ota}
  et~al.}{2008}]{2008ApJ...677...12O}
{Ota} K.,  et~al., 2008, \apj, 677, 12

\bibitem[\protect\citeauthoryear{{Ouchi} et~al.,}{{Ouchi}
  et~al.}{2010}]{2010ApJ...723..869O}
{Ouchi} M.,  et~al., 2010, \apj, 723, 869

\bibitem[\protect\citeauthoryear{{Paardekooper}, {Khochfar} \& {Dalla
  Vecchia}}{{Paardekooper} et~al.}{2013}]{2013MNRAS.429L..94P}
{Paardekooper} J.-P.,  {Khochfar} S.,    {Dalla Vecchia} C.,  2013, \mnras,
  429, L94

\bibitem[\protect\citeauthoryear{{Pandolfi}, {Ferrara}, {Choudhury},
  {Melchiorri} \& {Mitra}}{{Pandolfi} et~al.}{2011}]{2011PhRvD..84l3522P}
{Pandolfi} S.,  {Ferrara} A.,  {Choudhury} T.~R.,  {Melchiorri} A.,    {Mitra}
  S.,  2011, \prd, 84, 123522

\bibitem[\protect\citeauthoryear{{Pawlik}, {Schaye} \& {van
  Scherpenzeel}}{{Pawlik} et~al.}{2009}]{2009MNRAS.394.1812P}
{Pawlik} A.~H.,  {Schaye} J.,    {van Scherpenzeel} E.,  2009, \mnras, 394,
  1812

\bibitem[\protect\citeauthoryear{{Peng} et~al.,}{{Peng}
  et~al.}{2010}]{2010ApJ...721..193P}
{Peng} Y.-j.,  et~al., 2010, \apj, 721, 193

\bibitem[\protect\citeauthoryear{{Planck Collaboration} et~al.,}{{Planck
  Collaboration}  et~al.}{2015}]{2015arXiv150201589P}
{Planck Collaboration} et~al., 2015, arXiv:1502.01589

\bibitem[\protect\citeauthoryear{{Pritchard}, {Loeb} \& {Wyithe}}{{Pritchard}
  et~al.}{2010}]{2010MNRAS.408...57P}
{Pritchard} J.~R.,  {Loeb} A.,    {Wyithe} J.~S.~B.,  2010, \mnras, 408, 57

\bibitem[\protect\citeauthoryear{{Razoumov} \& {Sommer-Larsen}}{{Razoumov} \&
  {Sommer-Larsen}}{2010}]{2010ApJ...710.1239R}
{Razoumov} A.~O.,  {Sommer-Larsen} J.,  2010, \apj, 710, 1239

\bibitem[\protect\citeauthoryear{{Robertson}, {Ellis}, {Furlanetto} \&
  {Dunlop}}{{Robertson} et~al.}{2015}]{2015ApJ...802L..19R}
{Robertson} B.~E.,  {Ellis} R.~S.,  {Furlanetto} S.~R.,    {Dunlop} J.~S.,
  2015, \apjl, 802, L19

\bibitem[\protect\citeauthoryear{{Samui}, {Srianand} \& {Subramanian}}{{Samui}
  et~al.}{2007}]{2007MNRAS.377..285S}
{Samui} S.,  {Srianand} R.,    {Subramanian} K.,  2007, \mnras, 377, 285

\bibitem[\protect\citeauthoryear{{Schenker}, {Ellis}, {Konidaris} \&
  {Stark}}{{Schenker} et~al.}{2014}]{2014ApJ...795...20S}
{Schenker} M.~A.,  {Ellis} R.~S.,  {Konidaris} N.~P.,    {Stark} D.~P.,  2014,
  \apj, 795, 20

\bibitem[\protect\citeauthoryear{{Schneider}, {Salvaterra}, {Ferrara} \&
  {Ciardi}}{{Schneider} et~al.}{2006}]{2006MNRAS.369..825S}
{Schneider} R.,  {Salvaterra} R.,  {Ferrara} A.,    {Ciardi} B.,  2006, \mnras,
  369, 825

\bibitem[\protect\citeauthoryear{{Schroeder}, {Mesinger} \&
  {Haiman}}{{Schroeder} et~al.}{2013}]{2013MNRAS.428.3058S}
{Schroeder} J.,  {Mesinger} A.,    {Haiman} Z.,  2013, \mnras, 428, 3058

\bibitem[\protect\citeauthoryear{{Sobacchi} \& {Mesinger}}{{Sobacchi} \&
  {Mesinger}}{2013}]{2013MNRAS.432.3340S}
{Sobacchi} E.,  {Mesinger} A.,  2013, \mnras, 432, 3340

\bibitem[\protect\citeauthoryear{{Songaila} \& {Cowie}}{{Songaila} \&
  {Cowie}}{2010}]{2010ApJ...721.1448S}
{Songaila} A.,  {Cowie} L.~L.,  2010, \apj, 721, 1448

\bibitem[\protect\citeauthoryear{{Totani} et~al.,}{{Totani}
  et~al.}{2006}]{2006PASJ...58..485T}
{Totani} T.,  et~al., 2006, \pasj, 58, 485

\bibitem[\protect\citeauthoryear{{White}, {Becker}, {Fan} \& {Strauss}}{{White}
  et~al.}{2003}]{2003AJ....126....1W}
{White} R.~L.,  {Becker} R.~H.,  {Fan} X.,    {Strauss} M.~A.,  2003, \aj, 126,
  1

\bibitem[\protect\citeauthoryear{{Wood} \& {Loeb}}{{Wood} \&
  {Loeb}}{2000}]{2000ApJ...545...86W}
{Wood} K.,  {Loeb} A.,  2000, \apj, 545, 86

\bibitem[\protect\citeauthoryear{{Worseck} et~al.,}{{Worseck}
  et~al.}{2014}]{2014MNRAS.445.1745W}
{Worseck} G.,  et~al., 2014, \mnras, 445, 1745

\bibitem[\protect\citeauthoryear{{Wyithe} \& {Loeb}}{{Wyithe} \&
  {Loeb}}{2005}]{2005ApJ...625....1W}
{Wyithe} J.~S.~B.,  {Loeb} A.,  2005, \apj, 625, 1

\bibitem[\protect\citeauthoryear{{Yajima}, {Choi} \& {Nagamine}}{{Yajima}
  et~al.}{2011}]{2011MNRAS.412..411Y}
{Yajima} H.,  {Choi} J.-H.,    {Nagamine} K.,  2011, \mnras, 412, 411

\end{thebibliography}
\bibliographystyle{mn2e}

\end{document}